\documentclass[12pt]{article}

\usepackage[utf8]{inputenc} 
\usepackage[T1]{fontenc}    
\usepackage{hyperref}       
\usepackage{url}            
\usepackage{booktabs}       
\usepackage{amsfonts}       
\usepackage{nicefrac}       
\usepackage{microtype}      
\usepackage{lipsum}
\usepackage{fancyhdr}       
\usepackage{graphicx}       
\graphicspath{{media/}}     

\usepackage{pifont}
\title{Should we agree to disagree about Twitter's bot problem?
}

\author{
  Onur Varol \\
  Faculty of Engineering and Natural Sciences, Sabanci University \\
  Istanbul, Türkiye\\
  \texttt{onur.varol@sabanciuniv.edu} \\
}

\begin{document}
\maketitle

\begin{abstract}
Bots, simply defined as accounts controlled by automation, can be used as a weapon for online manipulation and pose a threat to the health of platforms. Researchers have studied online platforms to detect, estimate, and characterize bot accounts.
Concerns about the prevalence of bots were raised following Elon Musk's bid to acquire Twitter. Twitter's recent estimate that 5\% of monetizable daily active users being bot accounts raised questions about their methodology. 
This estimate is based on a specific number of active users and relies on Twitter's criteria for bot accounts. 
In this work, we want to stress that crucial questions need to be answered in order to make a proper estimation and compare different methodologies. 
We argue how assumptions on bot-likely behavior, the detection approach, and the population inspected can affect the estimation of the percentage of bots on Twitter. 
Finally, we emphasize the responsibility of platforms to be vigilant, transparent, and unbiased in dealing with threats that may affect their users.
\end{abstract}



\newpage
\section*{Introduction}

Social networking platforms provide the ability to communicate through a medium that hosts millions of accounts. Some of these accounts are partially or fully controlled by software to automate content creation and distribution, network structure, and the presentation of online personas~\cite{ferrara2016rise,cresci2020decade}. 
The existence of such accounts can be beneficial for certain use cases to foster efficient communication~\cite{monsted2017evidence,pennycook2020fighting}, affect segregation of user, and eliminate platform biases~\cite{wang2022information,freelon2022black}. However, current research focuses primarily on the impact of bots when used by malicious organisations to manipulate public opinion~\cite{shao2018spread,vosoughi2018spread,lazer2018science,varol2018deception,starbird2019disinformation}.

Research on bots stems from the risks that exist, especially in sensitive areas such as politics and health. Researchers have developed machine learning systems to detect these automated entities~\cite{chavoshi2016debot,cresci2016dna,sayyadiharikandeh2020detection,ng2022botbuster} and study their role in information spread~\cite{varol2020journalists,shao2018spread,vosoughi2018spread}.
Our 2017 research is still the largest analysis of bot prevalence on Twitter~\cite{varol2017online}. We analyzed over 14 million active accounts that use English as their primary language. We identified these accounts based on two criteria: i) more than 200 tweets in total and ii) more than 90 tweets in the last three months. By collecting information required by a social bot detection system called Botometer, we estimated 9\%-15\% of accounts that exhibit bot-like behavior.

\begin{figure}[t!]
    \centering
    \includegraphics[width=\linewidth]{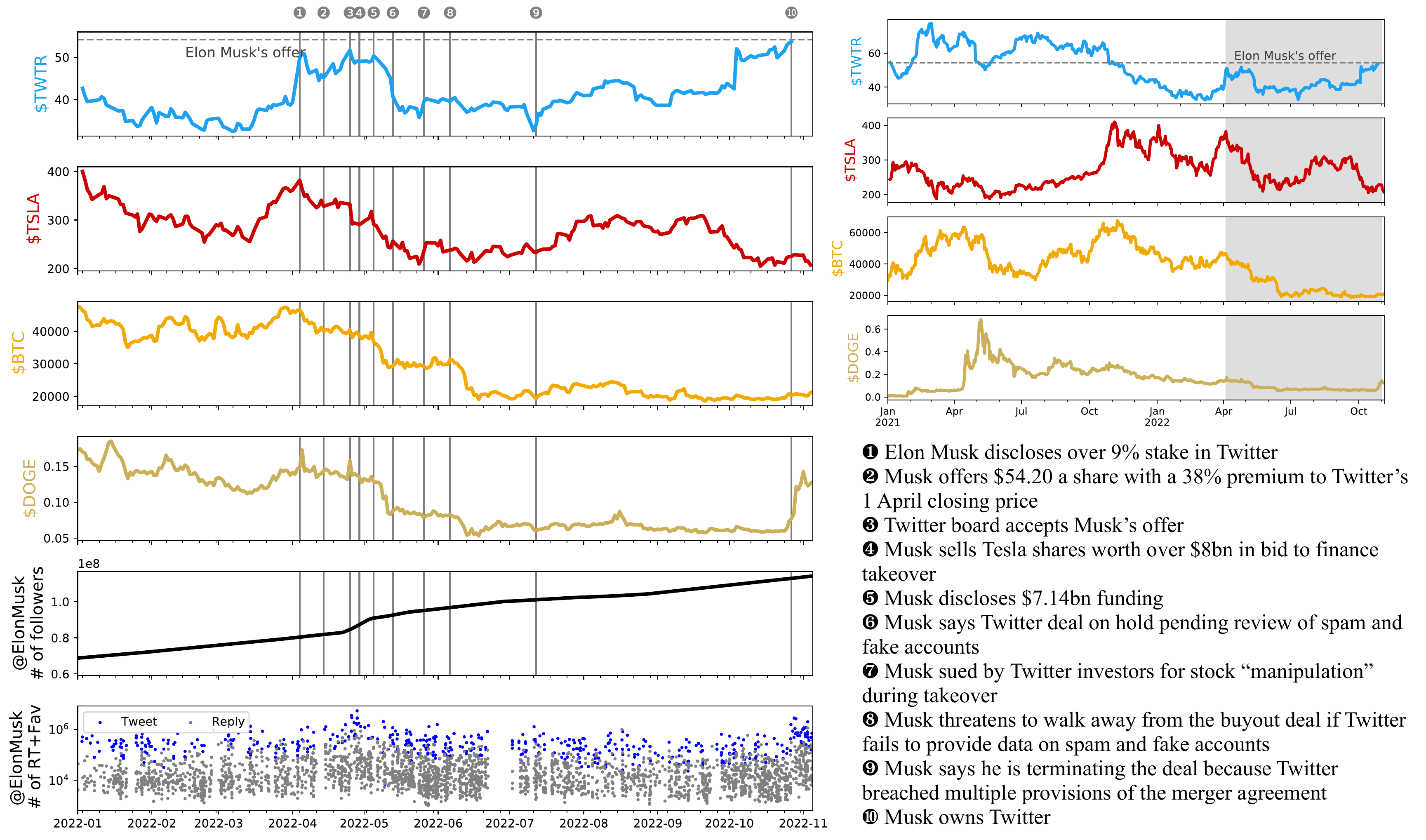}
    \caption{Timeline of Twitter v. Musk in relation to financial and online metrics. We present how stock prices for Twitter (\texttt{\$TWTR}), Tesla (\texttt{\$TSLA}), and two cryptocurrencies Bitcoin and DogeCoin (\texttt{\$BTC}, \texttt{\$DOGE}) changes. Follower counts and engagement metrics for Elon Musk's Twitter account (\texttt{@ElonMusk}) also presented for the last year.}
    \label{fig:timeline}
\end{figure}

Bot population estimation has implications not only for the field of online manipulation, but also for the valuation of companies, as this can be an important metric for investors. 
The current discussion about the legal battle between Twitter and Musk revolves around Twitter bots and their proliferation on the network. In Fig.~\ref{fig:timeline}, we present a timeline of events and show how some of the financial and online metrics change in relation to these events. 
The first notable event is Musk's announcement of a 9.2\% stake in Twitter on April 4. 
He later made an offer to buy Twitter, offering \$54.20 per share. Since the offer and Musk's official filing with SEC, we have observed a series of conversations between Elon Musk and former and current Twitter employees. Currently, Musk claims that Twitter's reports about the bot population on the platform are misleading, and he refused to proceed with the transaction.
After several months of debates, Musk purchased Twitter on October 27, 2022.

We can observe how the current debate affects both stock prices and online activity. For example, Musk's follower count increased much faster in May 2022, when he was seeking funding to buy Twitter. His tweets can also affect stock prices and investor behavior. His concern about bots also coincides with the date that Twitter shares lost significant value. Significant changes in stocks and crypto markets also occur around key events related to the deal, as seen in Fig.~\ref{fig:timeline}. Once the Twitter deal was closed by Musk's purchase of Twitter, there was a significant increase in Doge coin price.

Elon Musk's online popularity is reflected in the engagement of his content. Analyzing his recent tweets (retweets, replies and quotes from the last 3,200 tweets excluded), we found that the average engagement is over 240k. His 10 most popular pieces of content before the deal closed past year are all related to Twitter's acquisition. 
Some of his tweets are jokes about acquisitions. For instance on April 28, he tweeted, ``Next I'm buying Coca-Cola to put cocaine back in\footnote{https://twitter.com/elonmusk/status/1519480761749016577}'' which resulted in more than 5.5 million engagements. His second and third tweets with the highest engagement rates were about free speech and reached more than 6.5 million engagements.\footnote{https://twitter.com/elonmusk/status/1518623997054918657}\footnote{https://twitter.com/elonmusk/status/1518677066325053441}


To make a fair assessment about Elon Musk's claim about bots, we must first critically address three crucial questions: i) What is a bot? ii) How many accounts exists? and iii) How to identify potential risks?

\section*{Q1: What is a bot account?}

One of the most important issues is to find an appropriate and commonly agreed definition for bot accounts. To make comparisons, we must first need to establish on definitions. Social media companies, researchers, and journalists may have overlapping but slightly different definitions for bot accounts. These slight differences can lead to disagreements about the extent of the bot problem on these platforms.

\subsection*{How Twitter defines bots?}

Since Twitter is at the center of all debates about bots, their definition of bot is important. The company has stated in the past that it is actively fighting online manipulation, coordinated activity, and malicious bot accounts.\footnote{https://help.twitter.com/en/resources/addressing-misleading-info} 
They are also proactive in initially vetting new accounts, and recently Twitter CEO Parag Agrawal stated that Twitter suspends over half a million accounts a day before they impact the platform.\footnote{https://twitter.com/paraga/status/1526237583058952192} 
These examples highlight the company's efforts to create a healthier platform for all users.


Although Twitter struggles with malicious automated accounts, they also help developers create automated accounts by providing an interface for marketing and other engagement activities.\footnote{https://developer.twitter.com/en/docs/tutorials/how-to-create-a-twitter-bot-with-twitter-api-v2} Twitter has also taken steps to collect data on self-identified bots. The activities and behaviors of these accounts can be classified as automated/bot activities by detection tools and recognized as bot accounts by humans.

Twitter's efforts to identify accounts that \textit{cause harm} might overshadow accounts that are compromised by 3rd-party applications, created as dormant account to be used in coordinated activities, and used as deceptive in the context that is not prioritized by the company. 

\subsection*{How bot detection systems define bots?}

Bot detection systems focus on detecting irregularities in account behavior. One of the most popular tools, Botometer, defines bots as a social media account that is at least partially controlled by software~\cite{ferrara2016rise,davis2016botornot,varol2017online,sayyadiharikandeh2020detection}. This system extracts signals from profile, content, temporal, and network information to provide continuous score of bot-likeliness~\cite{varol2018feature,varol2017online}. In the latest version~\cite{sayyadiharikandeh2020detection}, researchers have developed a system that evaluates each account for different scenarios such as spammers, fake followers, financial, etc. to cover different bot behaviors in the system.

Another popular tool, BotSentinel, defines inauthentic accounts as ``nefarious individuals pretending to be something they are not to deceive their followers and audience, or automated accounts (bots) developed to behave as humanly as possible with the intent to deceive.''\footnote{https://help.botsentinel.com/support/solutions/articles/64000108232-what-are-inauthentic-accounts-} In this system, automation is part of the equation, and they also focus on account intent. They track ``disruptive'' accounts that frequently harass other accounts and use offensive language. Similarly, ``problematic'' accounts that frequently target other accounts and frequently use malicious tactics to harass their targets are evaluated by this system.

Research activities and tools developed for bot detection focus on behavioral features to isolate organic behaviors from others. Current approaches detect account-level anomalies; however, recent efforts have also focused on identifying coordinated activity as bot-like behavior has become increasingly difficult to detect in recent years.

\subsection*{How journalists talk about bots?}

Especially after the current events, journalists have turned to researchers for an expert opinion on Twitter bots. Because journalists play an important role in public understanding of the issue, how they convey the consequences of bots and define automated activity matters.

One of the most comprehensive research on bots conducted by PEW research center~\cite{wojcik2018bots,stocking2018social}. The researchers analyzed over 1.2 million links shared on Twitter and identified the most popular links they examined. They found that bots were responsible for 66\% of tweeted links to sites focused on news and current events and 89\% of links to popular news aggregation sites~\cite{wojcik2018bots}. In another study, a survey of 4,581 U.S. adults was conducted to determine public perceptions and concerns about bots. About two-thirds of this representative sample have heard of social media bots, and of those, 66\% believe social media bots have a mostly negative impact on how well they are informed about current events, while only 11\% believe they have a mostly positive impact~\cite{stocking2018social}.

Journalists were among the first professions to adopt the use of social media for their work~\cite{mullin2015report}. They enrolled for Twitter's verification platform, use Periscope for live streaming, and share the breaking news on the platform while engaging with their followers as digital journalism rapidly grows and changes journalism practices~\cite{brems2017personal,molyneux2018journalists,varol2020journalists}. 

Journalists' selection of newsworthy material influences how automated accounts are presented. Aside from technology news, bots are usually discussed in a political context, focusing on the negative aspects of their use. In this context, bots usually spread misinformation and manipulate public opinion, so journalists tend to portray bots as malicious automated accounts.


Overall, it should be noted that most of these definitions are slightly biased towards bots being malicious entities that are controlled for nefarious activities. However, bots can also be used to share useful information and news~\cite{haustein2016tweets,lokot2016news}; as well as collaborate with accounts to provide useful services~\cite{smith2022artbhot,brachten2018threat,monsted2017evidence,deshpande2021self,savage2016botivist}. 

To answer the question ``What is a bot?'' we should focus on automated activities and coordination between accounts. Since the boundaries between automated and organic accounts are not clear, we should consider the bot-likeliness as a spectrum. Discussions of behavioral intentions should be considered as additional and nearly orthogonal dimensions. We will also argue that vulnerable accounts can be converted into automated accounts when 3rd-party apps are compromised or dormant accounts are used for coordinated activities.

\section*{Q2: How many accounts exist on Twitter?}

The answer to this question may change depending on the methodology used to identify and count each follower. Researchers who have API access can capture unique users using two main mechanisms:

\textbf{Tweet activity}: Twitter offers API endpoints to provide public tweets. People with developer accounts can stream about 1\% of all public tweets. Accounts with an elevated access can retrieve nearly 10\% of public tweets. Using this streamed data, one can analyze all tweets and collect unique users to count the number of social media accounts. This approach can be problematic because i) there may be accounts that never post or are in a dormant state, ii) Twitter may delete or suspend some accounts since the last observed tweet, and iii) the activities of dormant accounts can be controlled by 3rd-party applications on the platform.

\textbf{Network connectivity}: Even though it is tedious, one can collect the friends and followers of each account to exhaustively discover underlying social network. Since there may be some disconnected components, the initial seed selection for collection is strategically important. 
Since deleted and suspended accounts also lose their social ties, this approach can provide an accurate picture of accounts that have at least one friend or follower. There may be coordinated bot armies that never tweet and only follow each other. These accounts are virtually invisible to all methods available to users of the Twitter API.

These two approaches rely on the Twitter API, which is offered in various forms, such as a developer API, a data stream in the form of Gardenhose ($\sim10\%$), and a recently introduced Academic access. The limitations of these streams have been discussed in recent literature~\cite{morstatter2013sample,pfeffer2018tampering,pfeffer2022sample}. Research efforts that aim to count the number of users on platforms should be aware of these limitations and the sampling strategies engineered by Twitter. Samples collected via APIs are not statistically correct ``random samples'' but filtered tweets based on tweet IDs.

Twitter has records of every account ever created, making it the only source that can accurately answer the question, ``How many accounts are there on Twitter?'' 
Twitter's quarterly reports to SEC (U.S. Securities and Exchange Commission) provide information on monetizable daily active usage (mDAU). 
Twitter defines mDAU as individuals, organizations, or other accounts that have logged in or otherwise been authenticated and accessed Twitter through its website on a given day.
In its latest filing, Twitter reports 237.8 million average mDAU as of June 30, 2022 \cite{secfiling}. However, we could not find information on the total number of registered accounts on Twitter, whether or not they are monthly active during the current observation period. Twitter's mDAU metric is useful for evaluating active accounts within a given time period, but ignores dormant accounts that may tweet at any time in the future through automation.


\begin{figure}[t!]
    \centering
    \includegraphics[width=\linewidth]{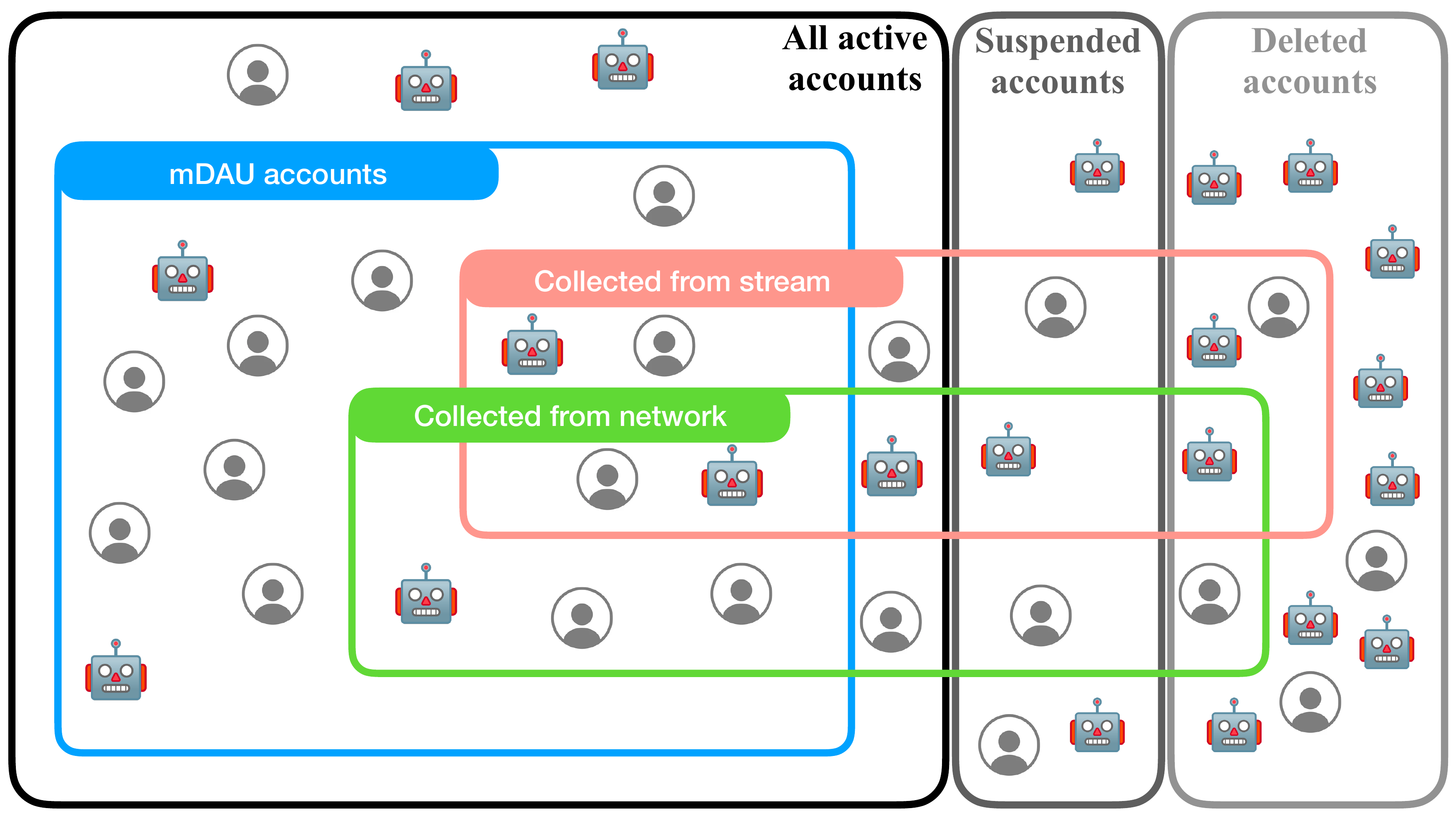}
    \caption{Properly estimating population of Twitter users require agreement on definitions of relevant accounts and data collection strategies.}
    \label{fig:population_estimate}
\end{figure}

Estimating the prevalence of bots in terms of percentiles requires not only the number of bot accounts on the platform, but also the total number of users. In Fig~\ref{fig:population_estimate}, we illustrate a space of social media accounts and how they can be observed through different collection methods. Above, we discussed the limitations of estimating the total number of accounts using different approaches. Twitter also only considers ``active'' accounts and acknowledges a potential underestimate in its reporting, as it reports in SEC filing as ``our estimation of false or spam accounts may not accurately represent the actual number of such accounts, and the actual number of false or spam accounts could be higher than we have estimated~\cite{secfiling}.'' In this analysis, Twitter estimates that 5\% of mDAU accounts are bot accounts; however, this analysis does not take into account accounts that are not active at the time of the analysis but pose a potential threat to the health of the platform.

To demonstrate our arguments in this section more clearly, we analyzed a publicly available dataset of Twitter from the Internet Web Archive.\footnote{\url{https://archive.org/search.php?query=collection\%3Atwitterstream\&sort=-publicdate}} We process tweets collected through the Twitter API and count i) the number of unique tweeting accounts, ii) the number of unique interacted accounts (mentioned, retweeted, quoted, and replied), and iii) the deletion statistics for those accounts.

\begin{figure}[t!]
    \centering
    \includegraphics[width=\linewidth]{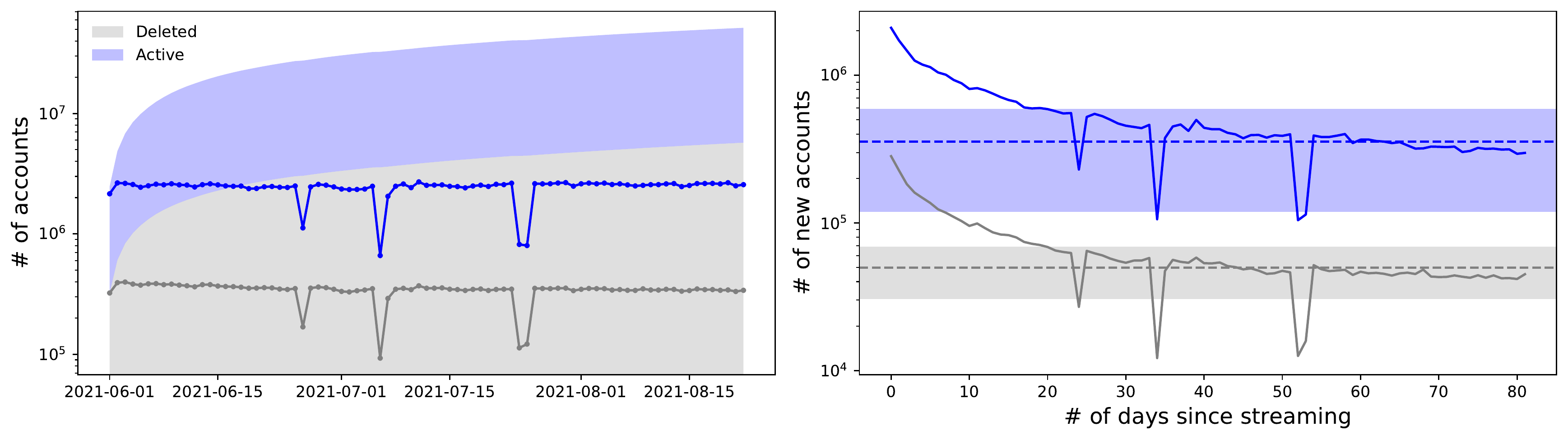}
    \caption{Analysis of tweet stream for gathering unique number of accounts that are currently active or deleted. On average there are 200,000 unique accounts that are still active (left). By observing data streams for more than 15 days, number of additional new accounts saturates (right).}
    \label{fig:archieve-tweetstats}
\end{figure}

To exhaustively account for the number of unique users on Twitter, we analyzed the daily tweets from the streams. This included not only the tweeting accounts, but also all other accounts that participated in these tweets by retweeting them, replying to them, or being mentioned in these tweets. In Fig.\ref{fig:archieve-tweetstats}(left), we show the cumulative number of unique accounts and also how many of them are still active in 2022. On average, there are two millions active accounts daily and about 500,000 accounts posting on these days are no longer active in the Twitter stream.

Our analysis of streaming data can also answer the question of how many days to observe the Twitter stream to capture as many users as possible. We tracked the cumulative number of unique users and calculated the derivative of the change for each day in Fig.\ref{fig:archieve-tweetstats}(right). In calculating the robust estimate for the number of novel accounts observed on each day, we identified about 350,000 accounts in the stream, with the first 15 days having significantly more observations because user activity rates vary and some users tweet less frequently than others. We can suggest that by observing Twitter stream for 15 days, one can capture significant portion of the active accounts on Twitter.

\begin{figure}[t!]
    \centering
    \includegraphics[width=\linewidth]{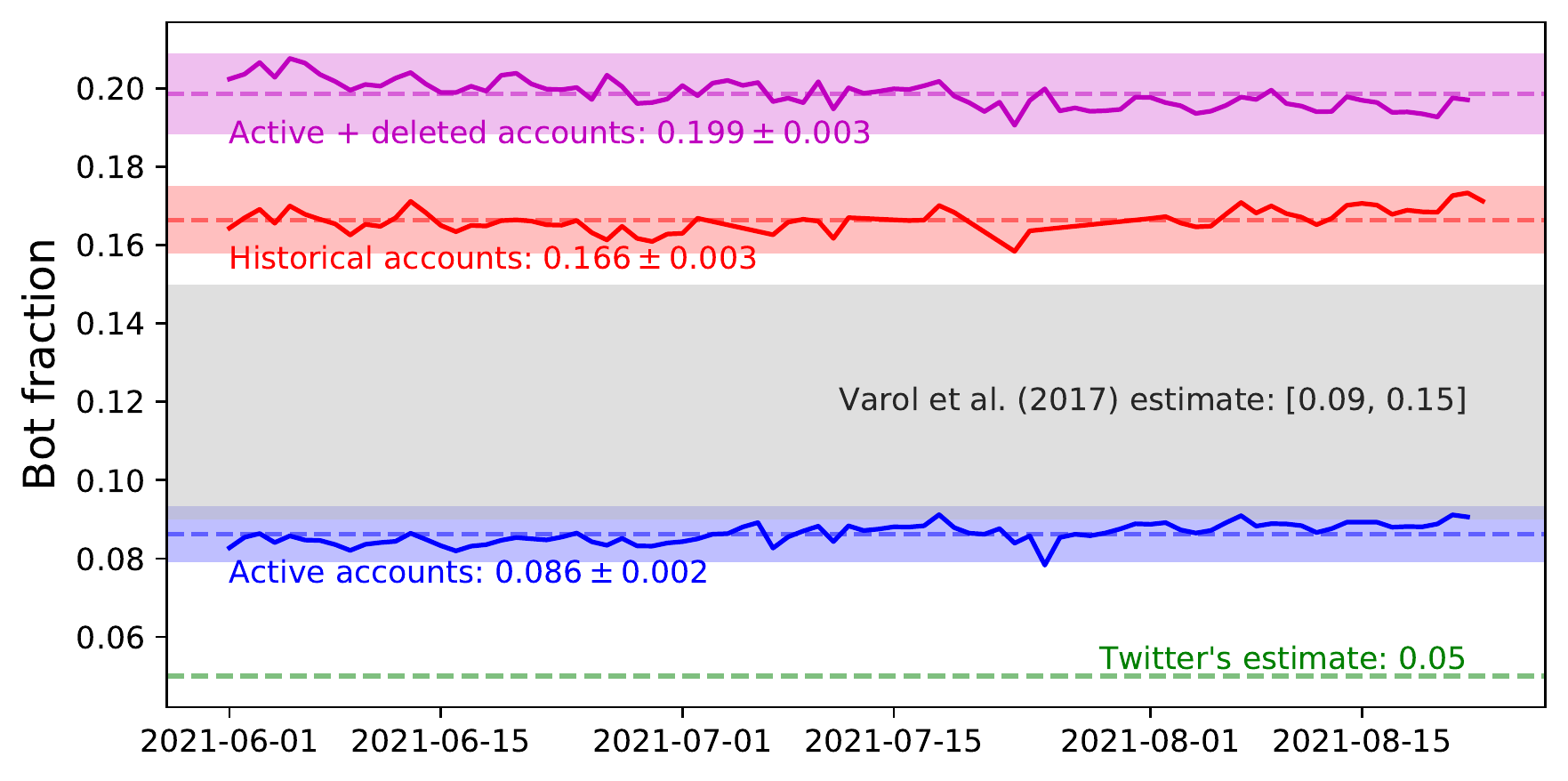}
    \caption{Estimating fraction of bots accounts over time. Active accounts (blue) consider only set off accounts still available on the platform. We can also use profile information of deleted accounts and conduct historical analysis (red) or we can make assumption that may lead overestimation by considering all deleted accounts being bot (pink). There are also two other estimates by Twitter (green) and Varol \textit{et al.} (2017).}
    \label{fig:archieve-bots}
\end{figure}

To offer estimates for bot-likely population, we utilize light-weighted model of Botometer, called BotometerLite, which can works with historical data and scalable since it only requires the profile meta-data~\cite{yang2020scalable}. In Fig.~\ref{fig:archieve-bots}, we compared different assumptions to build populations and we estimated bot scores and compared them available estimates for bot prevalence. 

When we analyzed the dataset, 11.3\% of accounts were deleted last year. We calculated bot scores for active accounts using current profile information. These active accounts have 8.6\% bot-likely population. However, this could be an underestimate since some of the bot accounts were detected and suspended by Twitter last year. If we assume that all deleted accounts are bots, this results in a higher and likely overestimated bot prevalence of about 20\%. To address this issue, we used historical data at the time of the tweet's creation to compute bot scores, and this approach resulted in an estimate of 16.5\% of daily active accounts being bot-likely.

Twitter's mDAU population covers not only active tweeting users but also passive accounts that logins to platform for solely consuming content. This large pool of accounts might lead to lower estimate of 5\% indicated by Twitter. Our 2017 paper estimated bot prevalence in the platform ranging between 9\% and 15\%. This estimation still reasonable and aligns with the observations in this work and the deletion statistics in the past one year.

\section*{Q3: How to identify potential risk?}

Bot detection has been an active research area since 2010~\cite{cresci2020decade}.
Considering the developments in deep learning technologies, we can observe profile pictures created by deepfake technologies~\cite{fagni2021tweepfake,narayan2022desi} and realistic messages through conversational bots~\cite{ccetinkaya2020developing,jeong2019improving}. These technologies lead to the development of novel automated behaviors and enrich the ways of interacting with the platform. Current approaches focus on models that can identify different types of bots and consider the bot detection problem as a binary classification between human and bot accounts~\cite{sayyadiharikandeh2020detection,yang2020scalable}.

These developments compel the arms race between bot detection and bot developers. 
As the boundaries between automated behavior and organic activity become more blurred, we should find ways to detect user intent and identify early warning signals of coordinated behaviors. 

There are still gray areas that require further research and policy recommendations for social media companies. Although these questions are not trivial to answer, they pose a risk if malicious organizations exploit them.

\textbf{How to treat a suspicious but inactive account?} Although Twitter's mDAU reaches 230 million, some accounts on the platform may be inactive. Accounts with no posts or social network do not provide sufficient information to determine their intentions or potential for harm. However, there are companies that sell aged accounts, and these accounts can be repurposed for various campaigns. The right to be forgotten laws protect such accounts, as they periodically delete all posts to remove traces of their automated activities. The platforms nudge accounts and periodically suspend inactive accounts, but they rarely check to see if a particular account changes ownership. These accounts, for which there is no clear evidence that it is a bot account, are considered human.

\textbf{How to regulate $3^{rd}$-party apps?} Twitter and many other social media platforms offer developers the ability to create applications that can automate tasks such as content creation, social interactions and private messaging. These applications manages permissions for thousands of users, and once compromised, they can be a tool for coordinated activity that controls content creation and dissemination. In 2017, there was a coordinated attack in which political messages were posted from thousands of accounts, some of which are popular accounts such as Forbes and BBC America. It is alleged that the Twitter Counter application was compromised and all accounts that provide their permissions to this application were used for the attack.

\textbf{Should Twitter act in benign looking bot armies?} Intentions are important in Twitter's definitions of being a bot accounts. The health of the platform takes precedence when it comes to suspending accounts from the platform, but the existence of coordinated bot armies still poses a problem because they can be used for coordinated campaigns. The researchers identified bot armies that simply post quotes from Star Wars~\cite{echeverria2017discovery}. They found that most of these accounts were active and systematically created on a large scale.

Twitter collects valuable information about how accounts access the platform, such as IP addresses and timestamps, as well as their latent interaction with content, such as impressions and clicks. These can be used to identify automated behavior, as manipulating them is more challenging than controlling their API. That said, not all bot-like behaviors are problematic, and Twitter supports developers in creating their own bots to automate certain activities.

\section*{Conclusion}

The activities of bots are an important problem for maintaining the health of the platform. Social media companies use various mechanisms to address concerns about the use of automation for spamming or malicious intent. These concerns were also raised when Elon Musk made bidding to acquire Twitter. The debate between Musk and Twitter is not an easy one to resolve, because the reported numbers for bot accounts, the methods used for bot detection, and the user populations studied vary quite significantly. 

We motivate this work to address important questions about platform-wide bot analysis and the challenges of estimating bot prevalence. ``What is a bot?'' is the first important question, as the use of automation and the intentions of accounts are difficult to disentangle. 
If only the bot accounts with bad intentions are considered, there is a risk to the health of the platform as the behavior of the accounts may change or they may be repurposed. 
Establishing what counts as a bot account will allow them to be identified and counted; however, there is more than one way to access user data on Twitter, which raises the second important question, ``Which accounts should be analyzed?'' Historical collection data from the stream may include deleted accounts and is not representative of dormant accounts, while network collection and mentioned accounts may capture them. Finally, there are gray areas where the platform can be proactive. We have to ask, ``How to identify potential risks?'' because vulnerabilities can be exploited and have consequences for platform.

Depending on the topic or type of social conversation, we can find different assessments of bots in the literature. For example, politics is one of the most sensitive topics, as the use of bots can manipulate political debates and the credibility of politicians~\cite{alsmadi2020many,bessi2016social,uyheng2020bot,stella2019influence,rossi2020detecting}. The use of bots was also investigated for the current public health threat that spread misinformation about vaccines during the COVID -19 pandemic~\cite{teng2022characterizing,yang2020prevalence,ferrara2020covid}. Because bot prevalence can depend on context and risk of manipulation, it is important to compare it in the same context. The BotAmp\footnote{https://botometer.osome.iu.edu/botamp/} application uses BotometerLite to provide a scalable tool for comparing different sets of online conversations~\cite{yang2020scalable}.

\begin{figure}[t!]
    \centering
    \includegraphics[width=\linewidth]{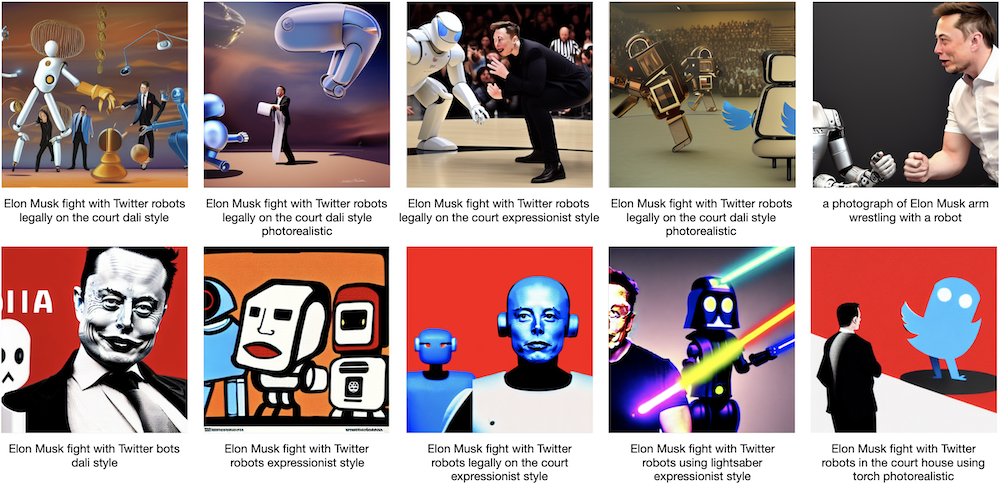}
    \caption{Text-to-image model, called Stable Diffusion, used with different textual prompts to create images.}
    \label{fig:ai-image}
\end{figure}

The development of deep-learning technologies and advances in text generation, deepfakes, and behavior modeling will pose new challenges to our fight against online manipulation. Although these technologies show promise (see Fig~\ref{fig:ai-image} for images created using StableDiffusion\footnote{\url{https://huggingface.co/spaces/stabilityai/stable-diffusion}}), there is still some progress needs to be done for becoming a real threat. In the meantime, the development of AI approaches to automated account detection and coordinated campaigns will arm users to fight malicious activity, and these tools will be a valuable asset to them~\cite{yang2019arming}.

Despite technology companies' efforts to fight misinformation and the use of automation to manipulate public opinion, users should also take responsibility and prioritize learning media literacy. 
Community-driven efforts are valuable to engage users in the process.
Twitter announced Birdwatch, a community-driven approach to misinformation, as a pilot for users from the United States in early 2021.\footnote{\url{https://blog.twitter.com/en_us/topics/product/2021/introducing-birdwatch-a-community-based-approach-to-misinformation}} This platform allows approved users to flag suspicious content and provide additional notes. Although this system is a great initiative to combat misinformation, researchers point out pitfalls and opportunities to improve this community-driven mechanism~\cite{prollochs2022community,allen2022birds,yasseri2021can}. A similar approach can be introduced for account-level annotations, flagging bots, trolls, sockpuppets, etc. This would be a step toward creating more transparent data requested by researchers~\cite{pasquetto2020tackling}. Community-driven efforts to label accounts can be used to initiate challenges to detect social bots~\cite{subrahmanian2016darpa}.

\section*{Acknowledgments}
I thank the Indiana University OSoMe team for fruitful conversations over the years and Aziz Simsir for his guidance in learning details about acquisitions. This work was supported in part by the TUBITAK Grant (121C220).

\newpage
\bibliographystyle{unsrt}  
\footnotesize
\bibliography{references}

\end{document}